\documentclass[twocolumn,showpacs,preprintnumbers,amsmath,amssymb]{revtex4}
\usepackage{dcolumn}
\usepackage{bm}
\usepackage{amssymb,amsmath, amsthm,epsfig,graphicx,mathrsfs,latexsym}

\newcommand{\da}{\partial}
\newcommand{\alf}{Alfv\'{e}n }

\begin{document}
\title{Mean shear flows generated by nonlinear resonant Alfv\'{e}n waves}
\author{Christopher TM Clack and Istvan Ballai}
\affiliation{Solar Physics and Space Plasma Research Centre
($SP^2RC$), Department of Applied Mathematics, University of
Sheffield, Hicks Building, Hounsfield Road, Sheffield, S3 7RH, U.K.}

\begin{abstract}
In the context of resonant absorption, nonlinearity has two
different manifestations. The first is the reduction in amplitude
of perturbations around the resonant point (wave energy
absorption). The second is the generation of mean shear flows
outside the dissipative layer surrounding the resonant point.
Ruderman \emph{et al.} [Phys. Plasmas {\bf 4}, 75 (1997)] studied
both these effects at the slow resonance in isotropic plasmas.
Clack \emph{et al.} [Astron. Astrophys. {\bf 494}, 317 (2009)]
investigated nonlinearity at the Alfv\'{e}n resonance, however,
they did not include the generation of mean shear flow. In this
present paper, we investigate the mean shear flow, analytically,
and study its properties. We find that the flow generated is
parallel to the magnetic surfaces and has a characteristic
velocity proportional to $\epsilon^{1/2}$, where $\epsilon$ is the
dimensionless amplitude of perturbations far away from the
resonance. This is, qualitatively, similar to the flow generated
at the slow resonance. The jumps in the derivatives of the
parallel and perpendicular components of mean shear flow across
the dissipative layer are derived. We estimate the generated mean
shear flow to be of the order of $10{\rm kms}^{-1}$ in both the
solar upper chromosphere and solar corona, however, this value
strongly depends on the choice of boundary conditions. It is
proposed that the generated mean shear flow can produce a
Kelvin--Helmholtz instability at the dissipative layer which can
create turbulent motions. This instability would be an additional
effect, as a Kelvin--Helmholtz instability may already exist due
to the velocity field of the resonant \alf waves. This flow can
also be superimposed onto existing large scale motions in the
solar upper atmosphere.
\end{abstract}
\pacs{52.25.Fi; 52.30.Cv; 52.35.-g; 52.35.Bj; 52.35.Mw}

\maketitle

\vspace{0.1cm}

\section{Introduction}

It has long been established, from observations, that the solar corona
is highly structured and inhomogeneous with temperatures of the order of $10^6\mbox{ K}$.
The solar corona is filled with a large number of discrete magnetic loops (coronal arcades) and there is
an abundance of observational evidence showing that magnetohydrodynamic (MHD) waves propagate in, and across, these loops (see, e.g. Refs. \citep{hassler1990}--\citep{banerjee2007}).
In order to sustain such high temperatures whilst combatting optically thin radiation and thermal conduction there
must exist some mechanism(s) acting as a source of steady heating.

The last few decades saw a multitude of models
proposed to tackle the complicated problem of coronal heating, such as heating by waves (e.g. resonant absorption,
phase mixing) or heating by magnetic relaxation (e.g. reconnection). There is an increasing
consensus that all these processes act simultaneously to different degrees of efficiency throughout the solar corona.
The agreement comes as observational data confirms the abundance of MHD wave propagation in the solar corona, along
with the complexity of the magnetic configuration enabling magnetic heating to occur (see, e.g. Refs. \citep{doschek1976}--\citep{acton1981}).

Resonant absorption of \alf waves in coronal loops was first
suggested by Ionson\citet{ionson1978} as a nonthermal heating
mechanism of the corona and since then resonant absorption has
been studied intensively by many authors in the linear regime for
a variety of applications (see, e.g. Refs.
\citep{kuperus1981}--\citep{poedts1990c}). A new understanding of
the process of resonant absorption became available after the
paper by Ruderman \emph{et al.}\citet{ruderman1997} which was the
first analytical study on the nonlinear aspect of resonant
absorption. Not only was it shown that resonant absorption of slow
waves was an inherently nonlinear phenomenon, they also showed
that a mean shear flow is generated outside the dissipative layer.
By their calculations, however, they still found the generated
flow was much too large compared to the observed velocities. On
the other hand, the authors did note that their results should be
used with caution in the solar atmosphere as some of their
assumptions were not fully realistic for that environment.

From the coronal heating point of view, the slow resonance studied by Ruderman \emph{et al.}\citet{ruderman1997} was not expected to contribute significantly as the energy stored in slow waves is much less than the required energy to compensate the losses and it is difficult for slow waves to reach the corona, as they become shocked as they climb due to density stratification. From this point of view, \alf waves and \alf resonance are of much greater interest as estimations show that the energy carried by \alf waves is much higher and they can reach the corona without significant damping. The validity of nonlinear resonant \alf waves (under coronal conditions) was studied in great detail by Clack \emph{et al.}\citet{clack2008b}, where they showed that in coronal plasmas the nonlinear addition to the result found in linear MHD is so small that the linear approach can be used with great accuracy. Essentially this investigation clarified the upper limit in which linear theory is applicable to resonant \alf waves.

Studies have been carried out to investigate the properties of shear flows generated by velocity field of \alf waves, however, nearly all of these have been numerical due to analytical complications when considering nonlinearity, turbulence and resonant absorption simultaneously. These studies have found that shear flows could give rise to a Kelvin--Helmholtz instability at the narrow dissipative layer (see, e.g. Ref. \citep{ofman1995}). This instability can drive turbulent motions and, in turn, locally enhance transport coefficients which can alter the efficiency of heating (see, e.g. Refs. \citep{ofman1995}--\citep{ofman1994}). None of these investigations have studied the generation of mean shear flow at the \alf resonance by nonlinear interactions. The generation of mean shear flow can supply additional shear enhancing turbulent motions.

Recent advancements in the understanding of nonlinear \alf resonance has inspired us to study of the generation of mean shear flows at the \alf resonance. In the present paper we will derive the equations describing the generated mean shear flow
outside the \alf dissipative layer and estimate the magnitude of the shear flow. We already know, from Ofman and Davila\citet{ofman1995}, that the plasma velocity at the dissipative layer may be reduced by the turbulent enhancement of the dissipative parameters, implying that for a given heating rate, the wave amplitude is reduced compared to the linear case. This means that any result we produce must be reduced when considered for the solar corona since turbulent motions are likely to be present and this will reduce the mean flow speed.

The paper is organised as follows. In the next section, we introduce the governing equations and discuss the main assumptions. In Sect. III we recall some previous results and derive the solution for the equation governing the mean shear flow outside the dissipative layer. In Sect. IV we derive the governing equations for the mean shear flow inside the dissipative layer. Section V gives the jumps in the derivatives of the
mean shear flow velocities across the dissipative layer and we estimate the magnitude of these jumps for conditions typical in the solar upper chromosphere and corona. In Sect. VI we draw our conclusions and discuss the results. Once the governing equations of the generated mean shear flow are found, the nonlinear theory of resonant \alf waves is complete.

\section{Governing equations and assumptions}

To mathematically study the mean shear flow generated at resonance we use the visco-resistive MHD
equations. In spite of the presence of dissipation, we use the adiabatic equation as an approximation to
the energy equation. Numerical studies by Poedts \emph{et al.}\citet{poedts1994} in linear MHD have shown that
dissipation due to viscosity and finite electrical conductivity in the energy equation does not, significantly,
alter the behaviour of resonant MHD waves in the driven problem.

When the product of the ion (electron) gyrofrequency, $\omega_{i(e)}$, and the ion (electron) collision time, $\tau_{i(e)}$,
is much greater than unity (as in the solar corona) the viscosity and finite electrical conductivity become anisotropic. The
parallel and perpendicular components of anisotropic finite electrical conductivity only differ by a factor of $2$, therefore,
we will consider only one of them without loss of generality. The anisotropic viscosity is given by Braginskii's viscosity tensor
(see, e.g. Braginskii \citep{braginskii1965}). The components of the viscosity tensor that remove the Alfv\'{e}n singularity
are the shear components, however these components (in their form) act as the isotropic viscosity. Indeed, Clack \emph{et al.}\citet{clack2008b} showed that one can interchange isotropic viscosity and shear viscosity when studying the Alfv\'{e}n resonance. Another possible additive to our model might be dispersive effects caused by Hall currents, however Clack \emph{et al.}\citet{clack2008b} showed that this effect does not alter the MHD waves in the vicinity of the \alf resonance.

The dynamics of waves in our model is described by the visco-resistive MHD equations
\begin{align}
&\frac{\da\overline{\rho}}{\da
t}+\nabla\cdot(\overline{\rho}\mathbf{v})=0,\mbox{
}\nabla\cdot\mathbf{B}=0,\label{eq:masscontinuityandsolenoid}\\
&\frac{\da\mathbf{v}}{\da
t}+(\mathbf{v}\cdot\nabla)\mathbf{v}=-\frac{1}{\overline{\rho}}\nabla\overline{P}+
\frac{1}{\mu_0\overline{\rho}}(\mathbf{B}\cdot\nabla)\mathbf{B}
+\overline{\nu}\nabla\cdot\mathbf{S}_1,\label{eq:momentum}\\
&\frac{\da\mathbf{B}}{\da
t}=\nabla\times(\mathbf{v}\times\mathbf{B})+\overline{\eta}\nabla^2\mathbf{B},\label{eq:induction}\\
&\frac{\da}{\da
t}\left(\frac{\overline{p}}{\overline{\rho}^{\gamma}}\right)+\mathbf{v}\cdot\nabla\left(\frac{\overline{p}}{\overline{\rho}^{\gamma}}\right)=0.\label{eq:adiabatic}
\end{align}
In Eqs.
(\ref{eq:masscontinuityandsolenoid})--(\ref{eq:adiabatic}),
$\mathbf{v}$ and $\mathbf{B}$ are the velocity and magnetic
induction vectors, $\overline{P}$ the total pressure, $\overline{p}$ the plasma pressure, $\overline{\rho}$ the
density, $\overline{\eta}$ the coefficient of magnetic
diffusivity, $\overline{\nu}=\overline{\eta}_1/\overline{\rho}$ the coefficient of shear viscosity over density, $\nabla\cdot\mathbf{S}_1$ the
shear component of Braginskii's viscosity tensor, $\gamma$ the adiabatic exponent, and $\mu_0$ the magnetic permeability of free space.

We adopt Cartesian coordinates $x, y, z$ and limit our analysis to a
static background equilibrium ($\mathbf{v}_0=0$). We assume that all
equilibrium quantities (terms with subscript `0') depend on $x$ only. The equilibrium magnetic
field, $\mathbf{B}_0$, is unidirectional and lies in the $yz$-plane.
The equilibrium quantities must satisfy the condition of total
pressure balance, $p_0+B_{0}^2/(2\mu_0)=\rm{constant}$.

For simplicity we assume that the perturbations of all quantities
are independent of $y$ ($\da/\da y =0$). We note that since the
magnetic field is not aligned with the $z$-axis, an Alfv\'{e}n resonance
can still exists. Even though the \alf resonance is governed by a linear equation, we must consider nonlinear effects to obtain the
second manifestation of nonlinearity - mean shear flows. In the linear theory of driven waves all perturbed
quantities oscillate with the same frequency, $\omega$, which means
that they can be Fourier-analysed and taken to be proportional to
$\exp(i[kz-\omega t])$. Solutions are sought in the form of
propagating waves and all perturbations in these solutions depend on
the combination $\theta=z-Vt$, rather than $z$ and $t$ separately,
with $V=\omega/k$. In the context of resonant absorption the phase velocity, $V$, must match the
projection of the Alfv\'{e}n velocity, $v_A$, onto the $z$-axis when
$x=x_a$ where $x_a$ is the resonant position. To define the resonant
position mathematically it is convenient to introduce the angle,
$\alpha$, between the $z$-axis and the direction of the equilibrium
magnetic field, so that the components of the equilibrium magnetic
field can be written as $B_{0y}=B_0\sin\alpha,\phantom{x}B_{0z}=B_0\cos\alpha$.
The definition of the resonant position can now be defined mathematically as
\begin{equation}\label{eq:resonantposition}
V=v_{A}\left(x_a\right)\cos\alpha,
\end{equation}
where $v_A$ is the \alf speed defined as $v_A=B_0/(\mu_0\rho_0)^{1/2}$.
In addition, we introduce the squares of the sound and cusp speeds as
$c^2_S=\gamma p_0/\rho_0$ and $c^2_T=c_S^2v_A^2/(c_S^2+v_A^2)$, respectively.
In what follows we can take $x_a=0$ without loss of generality. The
perturbations of the physical quantities are defined by
\begin{align}
&\overline{\rho}=\rho_0+\rho,\phantom{X} \overline{p}=p_0+p,\phantom{X}
\mathbf{B}=\mathbf{B}_0+\mathbf{b},\nonumber\\
&P=p+\frac{\mathbf{B}_0\cdot\mathbf{b}}{\mu_0}+\frac{\mathbf{b}^2}{2\mu_0},\label{eq:perturbations}
\end{align}
where $P$ is the perturbation of total pressure.

The dominant dynamics of resonant \alf waves, in linear MHD, resides
in the components of the perturbed magnetic field and velocity that
are perpendicular to the equilibrium magnetic field and to the
$x$-direction. This dominant behaviour is created by an $x^{-1}$
singularity in the spatial solution of these quantities at the \alf
resonance (see, e.g, Sakurai \emph{et al.} \citep{sakurai1991}, Goossens \emph{et al.} \citep{goossens1994}); these variables are known
as \emph{large variables}. All other variables are known as \emph{small variables}.

To make the mathematical analysis more concise, we define the components of velocity and magnetic field
that are in the $yz$-plane and are either parallel or perpendicular to the equilibrium magnetic
field:
\begin{align}
&\left(
\begin{array}{cc}
v_{\parallel}\\
b_{\parallel}
\end{array}\right)=\left(
\begin{array}{cc}
v\phantom{X} w\\
b_y\phantom{X} b_z
\end{array}\right)
\left(\begin{array}{cc}
\sin\alpha\\
\cos\alpha
\end{array}\right),\nonumber\\
&\left(
\begin{array}{cc}
v_{\perp}\\
b_{\perp}
\end{array}\right)=\left(
\begin{array}{cc}
v\phantom{X} -w\\
b_y\phantom{X} -b_z
\end{array}\right)
\left(\begin{array}{cc}
\cos\alpha\\
\sin\alpha
\end{array}\right),
\end{align}
where $v$, $w$, $b_y$ and $b_z$ are the $y$- and $z$-components of
the velocity and perturbation of magnetic field, respectively.

Let us introduce the characteristic scale of inhomogeneity,
$l_{\rm inh}$. The classical viscous Reynolds number, $R_e$, and the
magnetic Reynolds number, $R_m$, are defined as
\begin{equation}\label{eq:reynoldsnumbers}
R_e=\frac{Vl_{\rm inh}}{\overline{\nu}},\phantom{x}R_m=\frac{Vl_{\rm inh}}{\overline{\eta}}.
\end{equation}
These two numbers determine the importance of viscosity and finite electrical conductivity. We introduce the total Reynolds number as
\begin{equation}\label{eq:totalreynolds}
\frac{1}{R}=\frac{1}{R_e}+\frac{1}{R_m}.
\end{equation}
The aim of this paper is to study the generation of a mean shear flow outside the dissipative layer due to the
nonlinear behaviour of driven resonant Alfv\'{e}n waves in the dissipative layer. We are not interested in
the effects of MHD waves that have large amplitude everywhere and require a nonlinear description in the whole
space. We focus on waves that
have small dimensionless amplitude $\epsilon\ll 1$ far away from the
ideal \alf resonant point.

In nonlinear theory, when studying resonant behaviour in the
dissipative layer, we must re-scale the dissipative coefficients
(see, e.g., Ruderman \emph{et al.} \citep{ruderman1997}, Clack \emph{et al.} \citep{clack2008b}, Ballai \emph{et al.} \citep{ballai1998a} and Clack and Ballai \citep{clack2008}) so that dissipation is of the same order as nonlinearity,
\begin{equation}\label{eq:reycoefscales}
\overline{\nu}=\epsilon^{3/2}\nu,\mbox{ }\overline{\eta}=\epsilon^{3/2}\eta.
\end{equation}
It is easy to estimate the nonlinearity parameter, which is the ratio of the largest nonlinear and dissipative terms
and is obtained to be $\epsilon R^{2/3}$ (see Clack \emph{et al.} \citep{clack2008b}).
This implies that if $\epsilon R^{2/3}\ll 1$ linear theory holds, whereas if $\epsilon R^{2/3}\gtrsim 1$
then nonlinearity is important and linear theory breaks down.

In linear theory all perturbed quantities are harmonic functions of $\theta$, therefore their mean values
over a period vanish. On the other hand, in nonlinear theory the perturbed variables can have nonzero mean
values as a result of nonlinear interaction of different harmonics. Let us introduce the mean value
of a function $f(\theta)$ over a period $L$ as
\begin{equation}\label{eq:meanvalue}
\langle f\rangle=\frac{1}{L}\int_{0}^{L}f(\theta)\mbox{ }d\theta.
\end{equation}
It directly follows from Eq. (\ref{eq:masscontinuityandsolenoid}) that
\begin{equation}\label{eq:ubxzero}
\langle\overline{\rho}u\rangle=\langle b_x\rangle=0.
\end{equation}
We can always define the background state in such a way
that the mean values of density, pressure and magnetic field vanish:
\begin{equation}\label{eq:vanish1}
\langle\rho\rangle=\langle p\rangle=\langle b_{\parallel}\rangle=\langle b_{\perp}\rangle=0.
\end{equation}
This is not possible for the velocity, since we assume a static equilibrium. It is convenient, therefore,
to divide $v_{\parallel}$ and $v_{\perp}$ into mean and oscillatory parts. Using the Reynolds decomposition
we can write
\begin{align}
&U_{\parallel}=\langle v_{\parallel}\rangle,\quad U_{\perp}=\langle v_{\perp}\rangle,\nonumber\\
&\widetilde{v}_{\parallel}=v_{\parallel}-U_{\parallel},\quad \widetilde{v}_{\perp}=v_{\perp}-U_{\perp}.\label{eq:reynoldsdecomp}
\end{align}
The quantities $U_{\parallel}$ and $U_{\perp}$ describe the mean flow parallel to the magnetic
surfaces. The mean flow is generated by the nonlinear interaction of the harmonics in the
Fourier expansion of the perturbed quantities with respect to $\theta$.

We can rewrite Eqs. (\ref{eq:masscontinuityandsolenoid})--(\ref{eq:adiabatic}) in the scalar form as
\begin{equation}\label{eq:masscontinuity1}
V\frac{\da \rho}{\da\theta}-\frac{\da(\rho_0 u)}{\da
x}-\rho_0\frac{\da \widetilde{w}}{\da\theta}=\frac{\da(\rho u)}{\da
x}+\frac{\da(\rho w)}{\da\theta},
\end{equation}
\begin{multline}\label{eq:momentumx1}
\rho_0 V\frac{\da u}{\da\theta}-\frac{\da P}{\da
x}+\frac{B_0 \cos\alpha}{\mu_0}\frac{\da b_x}{\da\theta}=\\
\overline{\rho}\left(u\frac{\da u}{\da x}+w\frac{\da
w}{\da\theta}\right)
-\rho V\frac{\da u}{\da\theta}-\frac{b_x}{\mu_0}\frac{\da b_x}{\da x}\\
-\frac{b_z}{\mu_0}\frac{\da b_x}{\da\theta}
-\overline{\rho}\overline{\nu}(\nabla\cdot\mathbf{S}_1)_x,
\end{multline}
\begin{multline}\label{eq:momentumpara1}
\frac{\da}{\da\theta}\left(\rho_0 V
\widetilde{v}_{\parallel}-P\cos\alpha+\frac{B_0\cos\alpha}{\mu_0}b_{\parallel}\right)+\frac{b_x}{\mu_0}\frac{dB_0}{dx}=\\
\overline{\rho}\left(u\frac{\da v_{\parallel}}{\da x}+w\frac{\da
v_{\parallel}}{\da\theta}\right)
-\rho V\frac{\da
v_{\parallel}}{\da\theta}
-\frac{b_x}{\mu_0}\frac{\da b_{\parallel}}{\da x}\\
-\frac{b_z}{\mu_0}\frac{\da b_{\parallel}}{\da\theta}
-\overline{\rho}\overline{\nu}(\nabla\cdot\mathbf{S}_1)_{\parallel},
\end{multline}
\begin{multline}\label{eq:momentumperp1}
\frac{\da}{\da\theta}\left(\rho_0 V
\widetilde{v}_{\perp}+P\sin\alpha+\frac{B_0\cos\alpha}{\mu_0}b_{\perp}\right)=\\
\overline{\rho}\left(u\frac{\da v_{\perp}}{\da x}+w\frac{\da
v_{\perp}}{\da\theta}\right)
-\rho V\frac{\da
v_{\perp}}{\da\theta}
-\frac{b_x}{\mu_0}\frac{\da b_{\perp}}{\da x}\\
-\frac{b_z}{\mu_0}\frac{\da b_{\perp}}{\da\theta}
-\overline{\rho}\overline{\nu}(\nabla\cdot\mathbf{S}_1)_{\perp},
\end{multline}
\begin{equation}\label{eq:inductx1}
Vb_x+B_0 u\cos\alpha=wb_x-ub_z +\overline{\eta}\left(\frac{\da
b_x}{\da\theta}-\frac{\da b_z}{\da x}\right),
\end{equation}
\begin{multline}\label{eq:inductpara1}
\frac{\da}{\da\theta}\left(Vb_{\parallel}+B_0
\widetilde{v}_{\parallel}\cos\alpha\right)-\frac{\da(B_0 u)}{\da
x}-B_0\frac{\da w}{\da\theta}\\
=\frac{\da(ub_{\parallel})}{\da
x}+\frac{\da(wb_{\parallel})}{\da\theta}-b_x\frac{\da
v_{\parallel}}{\da x}-b_z\frac{\da v_{\parallel}}{\da\theta}
-\overline{\eta}\nabla^2 b_{\parallel},
\end{multline}
\begin{multline}\label{eq:inductperp1}
\frac{\da}{\da\theta}\left(Vb_{\perp}+B_0
\widetilde{v}_{\perp}\cos\alpha\right)=\frac{\da(ub_{\perp})}{\da
x}+\frac{\da(wb_{\perp})}{\da\theta}\\
-b_x\frac{\da v_\perp}{\da x}-b_z\frac{\da v_{\perp}}{\da\theta}
-\overline{\eta}\nabla^2 b_{\perp},
\end{multline}
\begin{multline}\label{eq:adiabatic1}
V\left(\frac{\da p}{\da\theta}-c_S^2\frac{\da \rho}{\da\theta}\right)-
u\left(\frac{dp_0}{dx}-c_S^2\frac{d\rho_0}{dx}\right)\\
=\frac{1}{\rho_0}\left\{V\left(\gamma p\frac{\da \rho}{\da \theta}-\rho\frac{\da p}{\da\theta}\right)
-w\left[\gamma\overline{p}\frac{\da\rho}{\da\theta}-\overline{\rho}\frac{\da p}{\da\theta}\right]\right.\\
\left.+u\left[\rho\frac{dp_0}{dx}-\gamma p\frac{d\rho_0}{dx}+\overline{\rho}\frac{\da p}{\da x}
-\gamma\overline{p}\frac{\da \rho}{\da x}\right]\right\}
\end{multline}
\begin{equation}\label{eq:totalpressure1}
P=p+\frac{1}{2\mu_0}\left(b_x^2+b_{\perp}^2+b_{\parallel}^2+2B_0
b_{\parallel}\right),
\end{equation}
The equations for $U_{\parallel}$ and $U_{\perp}$ are obtained by averaging Eqs. (\ref{eq:momentumpara1}) and (\ref{eq:momentumperp1}), respectively, and then dividing by $\overline{\rho}$,
\begin{multline}\label{eq:meanparallel}
\overline{\nu}\frac{d^2U_{\parallel}}{dx^2}=\left\langle u\frac{\da v_{\parallel}}{\da x}\right\rangle+\left\langle v_{\parallel}\frac{\da v_{\perp}}{\da\theta}\right\rangle\cos\alpha+\left\langle \frac{1}{\overline{\rho}}\frac{\da P}{\da\theta}\right\rangle\cos\alpha\\
-\frac{B_0\cos\alpha}{\mu_0}\left\langle\frac{1}{\overline{\rho}}\frac{\da b_{\parallel}}{\da\theta}\right\rangle
-\frac{1}{\mu_0}\left\langle\frac{b_x}{\overline{\rho}}\frac{\da b_{\parallel}}{\da x}\right\rangle
-\frac{\cos\alpha}{\mu_0}\left\langle\frac{b_{\parallel}}{\overline{\rho}}\frac{\da b_{\parallel}}{\da\theta}\right\rangle\\
+\frac{\sin\alpha}{\mu_0}\left\langle\frac{b_{\perp}}{\overline{\rho}}\frac{\da b_{\parallel}}{\da\theta}\right\rangle
-\frac{1}{\mu_0}\frac{dB_0}{dx}\left\langle\frac{b_x}{\overline{\rho}}\right\rangle,
\end{multline}
\begin{multline}\label{eq:meanperp}
\overline{\nu}\frac{d^2U_{\perp}}{dx^2}=\left\langle u\frac{\da v_{\perp}}{\da x}\right\rangle+\left\langle v_{\parallel}\frac{\da v_{\perp}}{\da\theta}\right\rangle\cos\alpha-\left\langle \frac{1}{\overline{\rho}}\frac{\da P}{\da\theta}\right\rangle\sin\alpha\\
-\frac{B_0\cos\alpha}{\mu_0}\left\langle\frac{1}{\overline{\rho}}\frac{\da b_{\perp}}{\da\theta}\right\rangle
-\frac{1}{\mu_0}\left\langle\frac{b_x}{\overline{\rho}}\frac{\da b_{\perp}}{\da x}\right\rangle\\
-\frac{\cos\alpha}{\mu_0}\left\langle\frac{b_{\parallel}}{\overline{\rho}}\frac{\da b_{\perp}}{\da\theta}\right\rangle
+\frac{\sin\alpha}{\mu_0}\left\langle\frac{b_{\perp}}{\overline{\rho}}\frac{\da b_{\perp}}{\da\theta}\right\rangle.
\end{multline}
Here $\nabla=(\da/\da x,0,\da/\da\theta)$, $u$ is the $x$-component
of the velocity, and
$\widetilde{w}=\widetilde{v}_{\parallel}\cos\alpha-\widetilde{v}_{\perp}\sin\alpha$ is the oscillatory part of $w$.

Equations (\ref{eq:masscontinuity1})--(\ref{eq:meanperp}) will be used in the
following sections in order to calculate the mean flow that is generated outside the
dissipative layer by the resonant waves.

\section{Outer solution}

To calculate the mean shear flow generated outside the dissipative layer we
have to recall some results found by Clack \emph{et al.}\citet{clack2008b}. They used
the method of simplified matched asymptotic expansions (see, e.g., Ballai \emph{et al.} \citep{ballai1998a} and Nayfeh \citep{nayfeh1981})
in order to derive the equation governing resonant Alfv\'{e}n waves in the dissipative layer. The method consists of
finding the so-called outer and inner expansions and matching these expansions in overlap regions around the dissipative layer. This nomenclature is well adopted for our scenario. The outer expansion corresponds to the solution outside the dissipative layer and the inner expansion
refers to the solution inside the dissipative layer. The solution in the outer region is represented by
asymptotic expansions of the form
\begin{equation}\label{eq:expansionform1}
f=\epsilon f^{(1)}+\epsilon^{3/2}f^{(2)}+\ldots,
\end{equation}
where $\epsilon$ is the dimensionless amplitude of perturbations far from the dissipative layer and $f$ represents
any perturbed quantity with the exception of $v_{\parallel}$ and $v_{\perp}$.
We shall show that resonant \alf waves create a shear flow with an amplitude proportional to $\epsilon^{1/2}$ outside the dissipative layer.
As a consequence, we expand $\widetilde{v}_{\parallel}$ and $\widetilde{v}_{\perp}$ in the form of Eq. (\ref{eq:expansionform1}),
and $U_{\parallel}$ and $U_{\perp}$ in the form
\begin{align}
&U_{\parallel}=\epsilon^{1/2}U^{(0)}_{\parallel}+\epsilon U^{(1)}_{\parallel}+\epsilon^{3/2}U^{(2)}_{\parallel}+\ldots,\nonumber\\
&U_{\perp}=\epsilon^{1/2}U^{(0)}_{\perp}+\epsilon U^{(1)}_{\perp}+\epsilon^{3/2}U^{(2)}_{\perp}+\ldots.\label{eq:meanexpansionoutside}
\end{align}

Clack \emph{et al.}\citet{clack2008b} found that the substitution of Eq. (\ref{eq:expansionform1}) into Eqs. (\ref{eq:masscontinuity1})--(\ref{eq:totalpressure1}) leads, in the first order approximation, to a system
of linear equations for the variables with the superscript `1'. All variables can be eliminated
in favour of $u^{(1)}$ and $P^{(1)}$, leading to the system
\begin{equation}\label{eq:linearequation}
V\frac{\da P^{(1)}}{\da\theta}=F\frac{\da u^{(1)}}{\da
x},\phantom{X} V\frac{\da P^{(1)}}{\da x}=\rho_0 D_{A}\frac{\da
u^{(1)}}{\da\theta},
\end{equation}
where
\begin{equation}\label{eq:F}
F=\frac{\rho_0 D_{A} D_{C}}{V^4-V^2\left(v_A^2+c_S^2\right)+v_A^2
c_S^2\cos^{2}\alpha},
\end{equation}
\begin{equation}\label{eq:A}
D_{A}=V^2-v_A^2\cos^{2}\alpha,\nonumber
\end{equation}
\begin{equation}\label{eq:C}
D_{C}=\left(v_A^2+c_S^2\right)\left(V^2-c_T^2\cos^{2}\alpha\right).
\end{equation}
The quantities $D_{A}$ and $D_{C}$ vanish at the \alf and slow resonant
positions, respectively. As a result these two positions are regular
singular points for the system (\ref{eq:linearequation}). The
remaining variables can be expressed in terms of $u^{(1)}$ and
$P^{(1)}$ as,
\begin{equation}\label{eq:relationlinear1}
\widetilde{v}_{\perp}^{(1)}=-\frac{V\sin\alpha}{\rho_0 D_{A}}P^{(1)}, \phantom{X}
\widetilde{v}_{\parallel}^{(1)}=\frac{Vc_S^2\cos\alpha}{\rho_0 D_{C}}P^{(1)},
\end{equation}
\begin{equation}\label{eq:relationlinear2}
b_x^{(1)}=-\frac{B_0\cos\alpha}{V}u^{(1)},\phantom{X}
b_{\perp}^{(1)}=\frac{B_0\cos\alpha\sin\alpha}{\rho_0 D_{A}}P^{(1)},
\end{equation}
\begin{equation}\label{eq:relationlinear3}
\frac{\da b_{\parallel}^{(1)}}{\da\theta}=\frac{B_0\left(
V^2-c_S^2\cos^{2}\alpha\right)}{\rho_0 D_{C}}\frac{\da
P^{(1)}}{\da\theta} +\frac{u^{(1)}}{V}\frac{dB_0}{dx},
\end{equation}
\begin{equation}\label{eq:relationlinear4}
\frac{\da p^{(1)}}{\da\theta}=\frac{V^2 c_S^2}{D_{C}}
\frac{\da P^{(1)}}{\da\theta}-\frac{u^{(1)}B_0}{\mu_0 V}\frac{dB_0}{dx},
\end{equation}
\begin{equation}\label{eq:relationlinear5}
\frac{\da \rho^{(1)}}{\da\theta}=\frac{V^2}{D_{C}}
\frac{\da P^{(1)}}{\da\theta}+\frac{u^{(1)}}{V}\frac{d\rho_0}{dx}.
\end{equation}
Since Eq. (\ref{eq:linearequation}) has regular singular points, the
solutions can be obtained in terms of Fr\"{o}benius series with
respect to $x$ (for details see, e.g., Clack \emph{et al.} \citep{clack2008b}, Ruderman \emph{et al.} \citep{ruderman1997}, Ballai \emph{et al.} \citep{ballai1998a}).
From Eqs. (\ref{eq:relationlinear1})--(\ref{eq:relationlinear5}), we see that
the quantity $\widetilde{v}_{\parallel}^{(1)}$ is regular, while all other
quantities are singular. The quantities $u^{(1)}$, $b_x^{(1)}$,
$b_{\parallel}^{(1)}$, $p^{(1)}$ and $\rho^{(1)}$ behave as
$\ln|x|$, while $\widetilde{v}_{\perp}^{(1)}$ and $b_{\perp}^{(1)}$ behave as
$x^{-1}$, so they are the most singular.

Carrying out calculations on Eqs. (\ref{eq:meanparallel}) and (\ref{eq:meanperp}), and
utilising Eqs. (\ref{eq:linearequation}), (\ref{eq:relationlinear1})--(\ref{eq:relationlinear5}), we find that in the first and second order approximations we have
\begin{align}\label{eq:flow1storder}
\frac{d^2 U^{(0)}_{\parallel}}{dx^2}=\frac{d^2 U^{(1)}_{\parallel}}{dx^2}=\frac{d^2 U^{(0)}_{\perp}}{dx^2}=\frac{d^2 U^{(1)}_{\perp}}{dx^2}=0.\\
\nonumber
\end{align}
The functions $U^{(0)}_{\parallel}(x)$, $U^{(0)}_{\perp}(x)$, $U^{(1)}_{\parallel}(x)$ and $U^{(1)}_{\perp}(x)$ are all continuous at $x=0$. Since
Eq. (\ref{eq:flow1storder}) implies that $U^{(0)}_{\parallel}(x)$, $U^{(0)}_{\perp}(x)$, $U^{(1)}_{\parallel}(x)$ and $U^{(1)}_{\perp}(x)$
are linear functions of $x$, we can include $U^{(1)}_{\parallel}(x)$ and $U^{(1)}_{\perp}(x)$ into $U^{(0)}_{\parallel}(x)$ and $U^{(0)}_{\perp}(x)$
and take $U^{(1)}_{\parallel}(x)=U^{(1)}_{\perp}(x)=0$, without loss of generality. We choose a mobile coordinate system such that
$U^{(0)}_{\parallel}(0)=U^{(0)}_{\perp}(0)=0$. Using this assumption in conjunction with Eq. (\ref{eq:flow1storder}) it follows that
\begin{equation}\label{eq:U0}
U^{(0)}_{\parallel}=V^{\pm}_{\parallel}x,\quad U^{(0)}_{\perp}=V^{\pm}_{\perp}x,
\end{equation}
where $V^{\pm}_{\parallel}$ and $V^{\pm}_{\perp}$ are constants and the superscripts `--' and `+' refer to $x<0$ and $x>0$, respectively.

Continuing in the same way we can find the solutions of the subsequent higher order approximations. At each step in the scheme
of approximations we obtain equations with the left-hand sides equal to the left-hand sides of the equations found in the
first order approximation. The right-hand sides of the equations are expressed in terms of variables of lower order approximations.
In carrying out this process we obtain that
\begin{equation}\label{eq:flownorder}
U^{(n)}_{\parallel}=O(x^{-n+1}),\quad U^{(n)}_{\perp}=O(x^{-n+1}),\quad n\geq 2.
\end{equation}
This means that the mean velocity starts to behave singularly from the third order approximation.

Taking into account Eqs. (\ref{eq:U0}) and (\ref{eq:flownorder}) we write the expansion for the
mean velocity in the form
\begin{equation}\label{eq:outsideflow}
U_{i}=\epsilon^{1/2}V_{i}x+\sum^{\infty}_{n=1}\epsilon^{(n+2)/2}V^{(n)}_{i}(x)x^{-n},
\end{equation}
where subscript `$i$' represents either the subscript `$\parallel$' or `$\perp$' and the functions
$V_{i}(x)$ and $V^{(n)}_{i}(x)$ have finite limits at $|x|\rightarrow 0$. The most important property
of expansion (\ref{eq:outsideflow}) is that the term of the lowest order approximation (proportional to $\epsilon^{1/2}$)
is very small inside the dissipative layer, but becomes large far away from the resonance. It is also interesting to note
that the remaining terms tend to zero far from resonance. This result
is in complete agreement with the studies by Ruderman \emph{et al.}\citet{ruderman1997} and Ballai \emph{et al.}\citet{ballai2000}, where they found the mean flow generated by resonant slow waves outside the dissipative layer to be proportional to $\epsilon^{1/2}$.

\section{Inner solution}

In this section we determine the inner expansion, which is the solution inside the dissipative layer.
The thickness of the dissipative layer is of the order of $l_{{\rm inh}}R^{-1/3}$, where $l_{{\rm inh}}$
is the characteristic scale of inhomogeneity. Since we assume that $R=\mathscr{O}(\epsilon^{-3/2})$ we have $l_{{\rm inh}}R^{-1/3}=\mathscr{O}(\epsilon^{1/2}l_{{\rm inh}})$. As a consequence it is convenient to introduce a stretching variable $\xi=\epsilon^{-1/2}x$
inside the dissipative layer.

We can rewrite Eqs. (\ref{eq:masscontinuity1})--(\ref{eq:totalpressure1}) using the stretching variable, however, to avoid repetition
of previous studies and for brevity, we only display the equations we use explicitly, which are the perpendicular component of momentum
and normal component of induction,
\begin{multline}\label{eq:momentumperp2}
\frac{1}{\overline{\rho}}\left[\epsilon^{1/2}\frac{\da
P}{\da\theta}\sin\alpha+\frac{b_x}{\mu_0}\frac{\da
b_{\perp}}{\da\xi}+\frac{\epsilon^{1/2}}{\mu_0}\left(B_0\cos\alpha+b_z\right)\frac{\da
b_{\perp}}{\da\theta}\right]\\
=-\epsilon^{1/2}\left(V-w\right)\frac{\da v_{\perp}}{\da\theta}+u\frac{\da
v_{\perp}}{\da\xi}-\epsilon\nu\left(\frac{\da^2 v_{\perp}}{\da\xi^2}+\epsilon\frac{\da^2v_{\perp}}{\da\theta^2}\right),
\end{multline}
\begin{multline}\label{eq:inductx2}
\epsilon^{1/2}\left(V-w\right)\frac{\da
b_x}{\da\theta}+\epsilon^{1/2}\left(B_0\cos\alpha+b_z\right)\frac{\da
u}{\da\theta}\\
+\epsilon\eta\left(\frac{\da^2b_x}{\da\xi^2}
+\epsilon\frac{\da^2b_x}{\da\theta^2}\right)=0,
\end{multline}
We use the stretched versions of Eqs. (\ref{eq:masscontinuity1})--(\ref{eq:totalpressure1}) to find the relationships between variables in each successive order of approximation (for full details we refer to Clack \emph{et al.}\citet{clack2008b}). Additionally, Eqs. (\ref{eq:momentumperp2}) and (\ref{eq:inductx2}) are used to help derive the equations governing the mean shear flow inside the dissipative layer.
Equations (\ref{eq:meanparallel}) and (\ref{eq:meanperp}) for the generated mean flow are transformed to
\begin{multline}\label{eq:meanparallel2}
\epsilon\nu\frac{d^2U_{\parallel}}{d\xi^2}=\left\langle u\frac{\da v_{\parallel}}{\da \xi}\right\rangle-\frac{1}{\mu_0}\left\langle\frac{b_x}{\overline{\rho}}\frac{\da b_{\parallel}}{\da \xi}\right\rangle\\
+\epsilon^{1/2}\left[\left\langle v_{\parallel}\frac{\da v_{\perp}}{\da\theta}\right\rangle\cos\alpha+\left\langle \frac{1}{\overline{\rho}}\frac{\da P}{\da\theta}\right\rangle\cos\alpha\right.\\
\left.-\frac{B_0\cos\alpha}{\mu_0}\left\langle\frac{1}{\overline{\rho}}\frac{\da b_{\parallel}}{\da\theta}\right\rangle
-\frac{\cos\alpha}{\mu_0}\left\langle\frac{b_{\parallel}}{\overline{\rho}}\frac{\da b_{\parallel}}{\da\theta}\right\rangle\right.\\
\left.+\frac{\sin\alpha}{\mu_0}\left\langle\frac{b_{\perp}}{\overline{\rho}}\frac{\da b_{\parallel}}{\da\theta}\right\rangle
-\frac{1}{\mu_0}\frac{dB_0}{dx}\left\langle\frac{b_x}{\overline{\rho}}\right\rangle\right],
\end{multline}
\begin{multline}\label{eq:meanperp2}
\epsilon\nu\frac{d^2U_{\perp}}{d\xi^2}=\left\langle u\frac{\da v_{\perp}}{\da \xi}\right\rangle-\frac{1}{\mu_0}\left\langle\frac{b_x}{\overline{\rho}}\frac{\da b_{\perp}}{\da \xi}\right\rangle\\
+\epsilon^{1/2}\left[\left\langle v_{\parallel}\frac{\da v_{\perp}}{\da\theta}\right\rangle\cos\alpha
-\frac{B_0\cos\alpha}{\mu_0}\left\langle\frac{1}{\overline{\rho}}\frac{\da b_{\perp}}{\da\theta}\right\rangle\right.\\
\left.-\left\langle\frac{1}{\overline{\rho}}\frac{\da P}{\da\theta}\right\rangle\sin\alpha-\frac{\cos\alpha}{\mu_0}\left\langle\frac{b_{\parallel}}{\overline{\rho}}\frac{\da b_{\perp}}{\da\theta}\right\rangle\right.\\
\left.+\frac{\sin\alpha}{\mu_0}\left\langle\frac{b_{\perp}}{\overline{\rho}}\frac{\da b_{\perp}}{\da\theta}\right\rangle\right].
\end{multline}

Following a similar procedure as introduced by Ruderman \emph{et al.}\citet{ruderman1997} we find that variables inside
the dissipative layer expand in a different way. The large variables ($\widetilde{v}_{\perp}$, $b_{\perp}$) have expansion of the form
\begin{equation}\label{eq:expansionlargeinside}
g=\epsilon^{1/2}g^{(1)}+\epsilon g^{(2)}+\ldots,
\end{equation}
whereas small variables ($P$, $p$, $\rho$, $u$, $\widetilde{v}_{\parallel}$, $b_x$, $b_{\parallel}$) expand as
\begin{equation}\label{eq:expansionsmallinside}
h=\epsilon h^{(1)}+\epsilon^{3/2}h^{(2)}+\ldots.
\end{equation}
The variables ${v}_{\parallel}$ and ${v}_{\perp}$ are expanded in series of the same form
as $\widetilde{v}_{\parallel}$ and $\widetilde{v}_{\perp}$, hence we expand $U_{\parallel}$ and $U_{\perp}$
as
\begin{equation}\label{eq:expansionmeaninside}
U_{\parallel}=\epsilon U^{(1)}_{\parallel}+\ldots,\quad U_{\perp}=\epsilon U^{(2)}_{\perp}+\ldots.
\end{equation}
We have chosen the superscripts in Eq. (\ref{eq:expansionmeaninside}) in such a way that
\begin{equation}\label{eq:reynoldsdecomp2}
v^{(n)}_{\parallel}=\widetilde{v}^{(n)}_{\parallel}+U^{(n)}_{\parallel},\quad v^{(n)}_{\perp}=\widetilde{v}^{(n)}_{\perp}+U^{(n)}_{\perp}.
\end{equation}

On substituting these expansions into the stretched versions of Eqs. (\ref{eq:masscontinuity1})--(\ref{eq:totalpressure1}) we find a linear homogeneous system of equations for the variables with superscript `1' in the first order approximation. This system parallels the results of linear theory (see, e.g., Clack \emph{et al.} \citep{clack2008b}). For the purposes of this paper, we are only interested in the expressions of variables found in terms of $u^{(1)}$, $\widetilde{v}^{(1)}_{\perp}$ and $P^{(1)}$:
 \begin{equation}
\widetilde{v}^{(1)}_{\parallel}=\frac{c_{Sa}^2}{v_{Aa}^2}\frac{\cos\alpha}{\rho_{0a}V}P^{(1)},\label{eq:1storder1}
\end{equation}
\begin{equation}\label{eq:1storder2}
b_{\perp}^{(1)}=-\frac{B_{0a}V}{v_{Aa}^2\cos\alpha}\widetilde{v}_{\perp}^{(1)},\mbox{ }
b_{x}^{(1)}=-\frac{B_{0a}\cos\alpha}{V}u^{(1)},
\end{equation}
\begin{equation}
\frac{\da b_{\parallel}^{(1)}}{\da\theta}=\frac{B_{0a}\left(v_{Aa}^2-c_{Sa}^2\right)}{\rho_{0a}v_{Aa}^4}
\frac{dP^{(1)}}{d\theta}+\frac{u^{(1)}}{V}\left(\frac{dB_0}{dx}\right)_a,\label{eq:1storder3}
\end{equation}
\begin{equation}
\frac{\da p^{(1)}}{\da\theta}=\frac{c_{Sa}^2}{v_{Aa}^2}
\frac{dP^{(1)}}{d\theta}-\frac{u^{(1)}}{V}\frac{B_{0a}}{\mu_0}\left(\frac{dB_0}{dx}\right)_a,\label{eq:1storder4}
\end{equation}
\begin{equation}\label{eq:1storder5}
\frac{\da\rho^{(1)}}{\da\theta}=\frac{1}{v_{Aa}^2}\frac{dP^{(1)}}{d\theta}+\frac{u^{(1)}}{V}\left(\frac{d\rho_0}{dx}\right)_a.
\end{equation}
In addition the equation that relates $u^{(1)}$ and
$\widetilde{v}_{\perp}^{(1)}$ is,
\begin{equation}\label{eq:reluvperp}
\frac{\da u^{(1)}}{\da\xi}=\frac{\da \widetilde{v}_{\perp}^{(1)}}{\da\theta}\sin\alpha.
\end{equation}
Here the subscript `$a$' means that the equilibrium quantities have been evaluated at the resonant position.
In the second order approximation, after eliminating all variables with superscript `1' in favour of $\widetilde{v}^{(1)}_{\perp}$ and $P^{(1)}$ using Eqs. (\ref{eq:1storder1})--(\ref{eq:reluvperp}), and satisfying the compatibility condition, we derive the equation governing resonant \alf waves inside the dissipative layer\citet{clack2008b}
\begin{equation}\label{eq:governinginside}
    \Delta_a\xi\frac{\da \widetilde{v}_{\perp}^{(1)}}{\da\theta} +
    V(\nu + \eta)\frac{\da^2 \widetilde{v}_\perp^{(1)}}{\da\xi^2} =
    -\frac{V\sin\alpha}{\rho_{0a}}\frac{dP^{(1)}}{d\theta},
\end{equation}
where $\Delta_a=-\left({dv_A^2}/{dx}\right)_a\cos^2\alpha$,
is the gradient of the square of the \alf speed.
This governing equation is linear despite it being derived by using the full nonlinear MHD equations, a result explained fully by Clack \emph{et al.}\citet{clack2008b}. Let us summarise their conclusions needed here for the sake of clarity. If all perturbations are expanded asymptotically, one can see that the second order terms describe magnetoacoustic modes and are, therefore, not resonant at the \alf resonance. In addition,
these terms act to cancel the small perturbations produced by the first order modes. This is in stark contrast to resonant slow waves,
where the second order terms act to enhance the perturbations from the first order.

Inserting the expansions (\ref{eq:expansionlargeinside})--(\ref{eq:expansionmeaninside}) into Eqs. (\ref{eq:meanparallel2})--(\ref{eq:meanperp2})
and collecting terms proportional to $\epsilon$, we have
\begin{multline}\label{eq:flowequationpara1}
\nu\frac{d^2U^{(1)}_{\parallel}}{d\xi^2}=\frac{\sin\alpha}{\mu_0\rho_{0a}}\left\langle b^{(1)}_{\perp}\frac{\da b^{(1)}_{\parallel}}{\da\theta}\right\rangle\\
+\sin\alpha\left\langle\widetilde{v}^{(1)}_{\parallel}\frac{\da \widetilde{v}^{(1)}_{\perp}}{\da\theta}\right\rangle
-\frac{1}{\mu_0\rho_{0a}}\left\langle b^{(1)}_x\frac{\da b^{(1)}_{\parallel}}{\da\xi}\right\rangle,
\end{multline}
\begin{multline}\label{eq:flowequationperp1}
\nu\frac{d^2U^{(2)}_{\perp}}{d\xi^2}=\left\langle u^{(1)}\frac{\da \widetilde{v}^{(2)}_{\perp}}{\da\xi}+u^{(2)}\frac{\da\widetilde{v}^{(1)}_{\perp}}{\da\xi}\right\rangle\\
-\frac{1}{\mu_0\rho_{0a}}\left\langle b^{(1)}_x\frac{\da b^{(2)}_{\perp}}{\da\xi}+b^{(2)}_x\frac{\da b^{(1)}_{\perp}}{\da\xi}\right\rangle
+\cos\alpha\left\langle\widetilde{v}^{(1)}_{\parallel}\frac{\da\widetilde{v}^{(1)}_{\perp}}{\da\theta}\right\rangle\\
-\frac{\cos\alpha}{\mu_0\rho_{0a}}\left\langle b^{(1)}_{\parallel}\frac{\da b^{(1)}_{\perp}}{\da\theta}\right\rangle
+\frac{B_{0a}\cos\alpha}{\mu_0\rho^{2}_{0a}}\left\langle\rho^{(1)}\frac{\da b^{(1)}_{\perp}}{\da\theta}\right\rangle\\
+\frac{\xi}{\mu_0\rho^2_{0a}}\left(\frac{d\rho_0}{dx}\right)_{a}\left\langle b^{(1)}_x\frac{\da b^{(1)}_{\perp}}{\da\xi}\right\rangle.
\end{multline}
In deriving Eqs. (\ref{eq:flowequationpara1}) and (\ref{eq:flowequationperp1}) we have utilized the fact that
\begin{equation}
\left\langle\frac{\da f}{\da\theta}\right\rangle=\left\langle f\frac{\da f}{\da\theta}\right\rangle=\left\langle\frac{\da g}{\da\theta}\right\rangle=\left\langle g\frac{\da g}{\da\theta}\right\rangle=0,\nonumber
\end{equation}
(which follows directly from Eq. (\ref{eq:meanvalue})) and Eqs. (\ref{eq:ubxzero}) and (\ref{eq:vanish1}). Now we use Eqs. (\ref{eq:1storder1})--(\ref{eq:reluvperp}) to eliminate terms on the right-hand sides of
Eqs. (\ref{eq:flowequationpara1}) and (\ref{eq:flowequationperp1}) in favour of $\widetilde{v}^{(1)}_{\perp}$ and $P^{(1)}$
to leave
\begin{equation}\label{eq:flowequationpara2}
\nu\frac{d^2U^{(1)}_{\parallel}}{d\xi^2}=-\frac{V\sin\alpha}{\rho_{0a}v^2_{Aa}\cos\alpha}
\left\langle\widetilde{v}^{(1)}_{\perp}\frac{dP^{(1)}}{d\theta}\right\rangle,
\end{equation}
\begin{multline}\label{eq:flowequationperp2}
\nu\frac{d^2U^{(2)}_{\perp}}{d\xi^2}=\left\langle u^{(1)}\frac{\da \widetilde{v}^{(2)}_{\perp}}{\da\xi}+u^{(2)}\frac{\da\widetilde{v}^{(1)}_{\perp}}{\da\xi}\right\rangle\\
-\frac{1}{\mu_0\rho_{0a}}\left\langle b^{(1)}_x\frac{\da b^{(2)}_{\perp}}{\da\xi}+b^{(2)}_x\frac{\da b^{(1)}_{\perp}}{\da\xi}\right\rangle\\
+\frac{1}{\rho_{0a}}\left(\frac{d\rho_0}{dx}\right)_{a}\left\langle u^{(1)}\frac{\da\left(\xi\widetilde{v}^{(1)}_{\perp}\right)}{\da\xi}\right\rangle\\
-\frac{1}{B_{0a}}\left(\frac{dB_0}{dx}\right)_{a}\left\langle u^{(1)}\widetilde{v}^{(1)}_{\perp}\right\rangle.
\end{multline}

Averaging the governing equation (\ref{eq:governinginside}) and substituting it into Eq. (\ref{eq:flowequationpara2}) we obtain
\begin{equation}\label{eq:flowequationpara3}
\frac{d^2U^{(1)}_{\parallel}}{d\xi^2}=-\frac{V\left(\nu+\eta\right)}{\nu v^2_{Aa}\cos\alpha}
\left\langle\widetilde{v}^{(1)}_{\perp}\frac{\da^2\widetilde{v}^{(1)}_{\perp}}{\da\xi^2}\right\rangle,
\end{equation}
which constitutes the equation that governs the generated mean flow inside the dissipative layer parallel
to the magnetic field lines. In order to derive the equivalent equation for the generated mean flow
perpendicular to the magnetic field lines we we note that Eqs. (\ref{eq:momentumperp2}) and (\ref{eq:inductx2}) in the second order approximation lead to
\begin{multline}\label{eq:momentumperp2ndorder}
\frac{\da\widetilde{v}^{(2)}_{\perp}}{\da\theta}+\frac{B_{0a}\cos\alpha}{\mu_0\rho_{0a}V}\frac{\da b^{(2)}_{\perp}}{\da\theta}=
-\frac{\sin\alpha}{\rho_{0a}V}\frac{dP^{(1)}}{d\theta}-\frac{\nu}{V}\frac{\da^2\widetilde{v}^{(1)}_{\perp}}{\da\xi^2}\\
+\left[\frac{B_{0a}}{\mu_0\rho_{0a}v^2_{Aa}}\left(\frac{dB_0}{dx}\right)_{a}-\frac{1}{\rho_{0a}}\left(\frac{d\rho_0}{dx}\right)_{a}\right]
\xi\frac{\da\widetilde{v}^{(1)}_{\perp}}{\da\theta},
\end{multline}
\begin{equation}\label{eq:inductionx2ndorder}
u^{(2)}+\frac{Vb^{(2)}_x}{B_{0a}\cos\alpha}=\frac{\eta\sin\alpha}{V}\frac{\da\widetilde{v}^{(1)}_{\perp}}{\da\xi}
-\frac{\xi u^{(1)}}{B_{0a}}\left(\frac{dB_0}{dx}\right)_{a}.
\end{equation}
Substituting Eq. (\ref{eq:governinginside}) into Eq. (\ref{eq:momentumperp2ndorder}) we obtain
\begin{multline}\label{eq:momentumperp3}
\frac{\da\widetilde{v}^{(2)}_{\perp}}{\da\xi}+\frac{B_{0a}\cos\alpha}{\mu_0\rho_{0a}V}\frac{\da b^{(2)}_{\perp}}{\da\xi}=
\left[\frac{\nu\Delta}{V^2(\nu+\eta)}\right.\\
\left.+\frac{1}{B_{0a}}\left(\frac{dB_0}{dx}\right)_a-\frac{1}{\rho_{0a}}\left(\frac{d\rho_0}{dx}\right)_a\right]\frac{\da\left(\xi\widetilde{v}^{(1)}_{\perp}\right)}{\da\xi}.
\end{multline}
On substitution of Eqs. (\ref{eq:inductionx2ndorder}) and (\ref{eq:momentumperp3}) into Eq. (\ref{eq:flowequationperp2})
we produce
\begin{multline}\label{eq:flowequationperp3}
\nu\frac{d^2U^{(2)}_{\perp}}{d\xi^2}=\frac{\eta\sin\alpha}{V}\left\langle\left(\frac{\da\widetilde{v}^{(1)}_{\perp}}{\da\xi}\right)^2\right\rangle\\
+\frac{\nu}{V^2(\nu+\eta)}\frac{\da}{\da\xi}\left(\Delta\xi\left\langle u^{(1)}\widetilde{v}^{(1)}_{\perp}\right\rangle\right).
\end{multline}
It follows directly from Eqs. (\ref{eq:reluvperp}) and (\ref{eq:governinginside}) that
\begin{multline}\label{eq:uvperpchange}
\frac{\Delta\xi}{V^2}\left\langle u^{(1)}\widetilde{v}^{(1)}_{\perp}\right\rangle=-\frac{\sin\alpha}{\rho_{0a}V}
\left\langle u^{(1)}P^{(1)}\right\rangle\\
-\frac{(\nu+\eta)}{V}u^{(1)}\int\left\langle\frac{\da^2\widetilde{v}^{(1)}_{\perp}}{\da\xi^2}\right\rangle d\theta
\end{multline}
and
\begin{equation}\label{eq:governingaveraged}
\left\langle P^{(1)}\frac{\da u^{(1)}}{\da\xi}\right\rangle=\rho_{0a}\left(\nu+\eta\right)\left\langle\widetilde{v}^{(1)}_{\perp}
\frac{\da^2\widetilde{v}^{(1)}_{\perp}}{\da\xi^2}\right\rangle.
\end{equation}
Using Eq. (\ref{eq:flowequationperp3}) along with Eqs. (\ref{eq:uvperpchange})--(\ref{eq:governingaveraged}) we find the
equation governing the generated mean shear flow inside the dissipative layer perpendicular to the magnetic field lines is
\begin{multline}\label{eq:flowequationperp4}
\frac{d^2U^{(2)}_{\perp}}{d\xi^2}=\frac{\eta\sin\alpha}{\nu V}\left\langle\left(\frac{\da\widetilde{v}^{(1)}_{\perp}}{\da\xi}\right)^2\right\rangle\\
-\frac{\sin\alpha}{V}\left\langle\widetilde{v}^{(1)}_{\perp}
\frac{\da^2\widetilde{v}^{(1)}_{\perp}}{\da\xi^2}\right\rangle
-\frac{1}{V}\left\langle\int\frac{\da^3\widetilde{v}^{(1)}_{\perp}}{\da\xi^3}d\theta\right\rangle.
\end{multline}

In the next section we will use Eqs. (\ref{eq:flowequationpara3}) and (\ref{eq:flowequationperp4}) along with the
compatibility condition to derive the jumps in the derivatives of the parallel and perpendicular components
of the mean velocity across the dissipative layer.

\section{Jump conditions}

To derive the jump in the derivative of the parallel component of the mean velocity across the dissipative layer
we remember from Eq. (\ref{eq:U0}) that $V^{\pm}_{\parallel}$ are constants and the superscripts `--' and `+' refer to $x<0$ and $x>0$, respectively. This means that
\begin{align}
V^{+}_{\parallel}-V^{-}_{\parallel}&=\lim_{\xi\rightarrow\infty}\frac{dU^{(1)}_{\parallel}}{d\xi}-\lim_{\xi\rightarrow -\infty}\frac{dU^{(1)}_{\parallel}}{d\xi}\nonumber\\
&=-\frac{V(\nu+\eta)}{\nu v^2_{Aa}\cos\alpha}\int^{\infty}_{-\infty}\left\langle\left(\frac{\da\widetilde{v}^{(1)}_{\perp}}{\da\xi}\right)^2\right\rangle d\xi.\label{eq:jumppara}
\end{align}
In a similar fashion, we find the jump in the derivative of the perpendicular component of the mean velocity across
the dissipative layer to be
\begin{align}
V^{+}_{\perp}-V^{-}_{\perp}&=\lim_{\xi\rightarrow\infty}\frac{dU^{(1)}_{\perp}}{d\xi}-\lim_{\xi\rightarrow -\infty}\frac{dU^{(1)}_{\perp}}{d\xi}\nonumber\\
&=\frac{(\nu+\eta)\sin\alpha}{\nu V}\int^{\infty}_{-\infty}\left\langle\left(\frac{\da\widetilde{v}^{(1)}_{\perp}}{\da\xi}\right)^2\right\rangle d\xi.\label{eq:jumpperp}
\end{align}

Outside the dissipative layer we have the following approximate equalities:
\begin{equation}\label{eq:outsideapprox}
U_{\parallel}\simeq\epsilon^{1/2}U^{(0)}_{\parallel}(x),\quad U_{\perp}\simeq\epsilon^{1/2}U^{(0)}_{\perp}(x).
\end{equation}
Let us introduce the new dimensionless variables
\begin{equation}
\sigma=\delta^{-1}_a x=\epsilon^{1/2}\delta^{-1}_a\xi,\quad q=\frac{\epsilon^{1/2}k\delta_a}{V}\widetilde{v}^{(1)}_{\perp},\nonumber
\end{equation}
where $\delta_a=\left[V\left(\overline{\nu}+\overline{\eta}\right)/\left(k|\Delta_a|\right)\right]^{1/3}$,
is the width of the dissipative layer and $k=2\pi/L$. Equations (\ref{eq:jumppara}) and (\ref{eq:jumpperp})
can be rewritten in these new variables as
\begin{equation}\label{eq:jumpparafinal}
\left[\frac{dU_{\parallel}}{dx}\right]=-\frac{|\Delta_a|\cos\alpha}{2\pi\overline{\nu}}\int^{L}_{0}d\theta
\int^{\infty}_{-\infty}\left(\frac{\da q}{\da\sigma}\right)^2d\sigma,
\end{equation}
\begin{equation}\label{eq:jumpperpfinal}
\left[\frac{dU_{\perp}}{dx}\right]=\frac{|\Delta_a|\sin\alpha}{2\pi\overline{\nu}}\int^{L}_{0}d\theta
\int^{\infty}_{-\infty}\left(\frac{\da q}{\da\sigma}\right)^2d\sigma.
\end{equation}
These equations are implicit connection formulae, and as such, they must be solved in conjunction with
Eq. (\ref{eq:governinginside}) (to determine the dimensionless variable $q$). We note that Eqs. (\ref{eq:jumpparafinal})
and (\ref{eq:jumpperpfinal}) imply that at $\alpha=0,\pi/2$ the jumps are zero, since at these values of $\alpha$
there is no \alf resonance present.

In fact, Eqs. (\ref{eq:jumpparafinal}) and
(\ref{eq:jumpperpfinal}) can be solved explicitly, since the
governing equation (\ref{eq:governinginside}) is linear. It has
been shown by, e.g. Goossens \emph{et al.}\citet{goossens1994} and
Erd\'{e}lyi\citet{erdelyi1997}, that the solution of the governing
equation can be found in terms of the so-called $F$ and $G$
functions. In order to find the solutions to Eqs.
(\ref{eq:jumpparafinal}) and (\ref{eq:jumpperpfinal}) we need to
find $\widetilde{v}^{(1)}_{\perp}$ in order to obtain $q$. In
cartesian coordinates, this was recently accomplished by
Ruderman\citet{ruderman2009} where he found that
\begin{equation}\label{eq:vperpf}
\widetilde{v}_{\perp}=\frac{iVP\sin\alpha}{\rho_{0a}\delta_a|\Delta_a|}F(\sigma),
\end{equation}
where
\begin{equation}\label{eq:functionF}
F(\sigma)=\int^{\infty}_{0}\exp{\left(i\phi\sigma{\rm sgn}(\Delta_a)-\phi^3/3\right)}\mbox{ }d\phi.
\end{equation}
{Equation (\ref{eq:vperpf}) is the cartesian version of Eq. (66)
found by Goossens \emph{et al.}\citet{goossens1994}.}

Substituting Eq. (\ref{eq:vperpf}) into Eqs. (\ref{eq:jumpparafinal}) and (\ref{eq:jumpperpfinal}) leads to
\begin{equation}\label{eq:jumpparaf}
\left[\frac{dU_{\parallel}}{dx}\right]=-A\cos\alpha\int^{L}_{0}P^2\mbox{ }d\theta \int^{\infty}_{-\infty} \left|\frac{dF}{d\sigma}\right|^2d\sigma,
\end{equation}
\begin{equation}\label{eq:jumpperpf}
\left[\frac{dU_{\perp}}{dx}\right]=A\sin\alpha\int^{L}_{0}P^2\mbox{ }d\theta \int^{\infty}_{-\infty} \left|\frac{dF}{d\sigma}\right|^2d\sigma,
\end{equation}
with $A=k^2\sin^2\alpha/\left(2\pi\overline{\nu}\rho^2_{0a}|\Delta_a|\right)$.
It is shown in the Appendix that
\begin{equation}\label{eq:fcalculated}
\int^{\infty}_{-\infty} \left|\frac{dF}{d\sigma}\right|^2d\sigma=\pi,
\end{equation}
implying that Eqs. (\ref{eq:jumpparaf}) and (\ref{eq:jumpperpf}) reduce to
\begin{equation}\label{eq:jumpparaf1}
\left[\frac{dU_{\parallel}}{dx}\right]=-\frac{k^2\sin^2\alpha\cos\alpha}
{2\overline{\nu}\rho^2_{0a}|\Delta_a|}\int^{L}_{0}P^2\mbox{ }d\theta,
\end{equation}
\begin{equation}\label{eq:jumpperpf1}
\left[\frac{dU_{\perp}}{dx}\right]=\frac{k^2\sin^3\alpha}{2\overline{\nu}\rho^2_{0a}|\Delta_a|}\int^{L}_{0}P^2\mbox{ }d\theta.
\end{equation}
Equations (\ref{eq:jumpparaf1}) and (\ref{eq:jumpperpf1}) are the
explicit connection formulae for the jumps in the derivatives of
the mean shear flow across the dissipative layer. They are
explicit because we are considering a driven problem, and hence
$P$ is assumed to be known.

If we take $\alpha=\pi/4$ we have the following approximation
\begin{equation}\label{eq:estimate}
\left[\frac{dU_{\parallel}}{dx}\right]=-\left[\frac{dU_{\perp}}{dx}\right]\simeq\frac{\epsilon^{1/2}V}{l_{{\rm inh}}},
\end{equation}
and these values can be seen as jumps in vorticity.
Here we have used the obvious estimates $|\Delta_a|=\mathscr{O}(V^2/l_{{\rm inh}})$, $P=\mathscr{O}(\epsilon V^2 k l_{{\rm inh}})$ and $\overline{\nu}=\mathscr{O}(\epsilon^{3/2}Vl_{{\rm inh}})$.
In order to find the profiles of the components of the generated mean shear flow we need to impose
boundary conditions far away from the dissipative layer. For example, if there are rigid walls at $x=\pm a$
where the condition of adhesion has to be satisfied, then the components of the generated mean flow take
the simple form
\begin{equation}\label{eq:simpleprofilepara}
U_{\parallel}=\left\{
\begin{array}{ll}
\displaystyle{\left[\frac{dU_{\parallel}}{dx}\right]\frac{x-a}{2},\quad x>0}, \\
\noalign{\vskip 0.2cm}
\displaystyle{-\left[\frac{dU_{\parallel}}{dx}\right]\frac{x+a}{2},\quad x<0},
\end{array} \right.
\end{equation}
\begin{equation}\label{eq:simpleprofileperp}
U_{\perp}=\left\{
\begin{array}{ll}
\displaystyle{\left[\frac{dU_{\perp}}{dx}\right]\frac{x-a}{2},\quad x>0}, \\
\noalign{\vskip 0.2cm}
\displaystyle{-\left[\frac{dU_{\perp}}{dx}\right]\frac{x+a}{2},\quad x<0}.
\end{array} \right.
\end{equation}

With the estimate given by Eq. (\ref{eq:estimate}) and the simple
mean flow profiles in  Eqs. (\ref{eq:simpleprofilepara}) and
(\ref{eq:simpleprofileperp}) we can calculate the expected mean
flow generated outside the dissipative layer in the solar upper
chromosphere and solar corona. For example, if the incoming wave
has a dimensionless amplitude of $\epsilon=\mathscr{O}(10^{-4})$,
then the predicted mean shear flow is of the order of $10{\rm
kms}^{-1}$ in both the upper chromosphere and corona. Here we have
assumed that the characteristic scale of inhomogeneity ($l_{\rm
inh}$) is $10^{2}{\rm m}$ in the upper chromosphere and $10^3{\rm
m}$ in the corona. This generated flow can be superimposed on
existing flow, so it is difficult to observe. These results should
be used with caution, when applied to the solar atmosphere. We
have assumed that outside the dissipative layer the plasma is
homogeneous and infinite, and clearly the solar atmosphere is
neither, so, since all flows are entirely governed by the boundary
conditions, the flow generated may be less than predicted or may
even be stopped entirely.

\section{Conclusions}

In the present paper we have completed the nonlinear theory of resonant \alf waves in dissipative layers
in a one-dimensional (1-D) planar geometry. Clack \emph{et al.} \citet{clack2008b} showed that even though
nonlinearity and dispersion are considered the equation governing resonant \alf waves in the dissipative layer
is always linear (provided $\epsilon\ll R^{-1/3}$). However, they neglected the second manifestation of nonlinearity at resonance;
the generation of mean shear flows outside the dissipative layer. This flow is produced by the nonlinear interaction
of harmonics inside the dissipative layer and may still exist even though the governing equation
inside the dissipative layer is linear.

We have shown that outside the dissipative layer a mean flow is generated parallel to the magnetic surfaces.
The flow has an amplitude proportional to $\epsilon^{1/2}$, and depends linearly on $x$. The derivatives of the velocity
of the generated mean flow have a nonzero jump across the dissipative layer determined by Eqs. (\ref{eq:jumpparafinal})
and (\ref{eq:jumpperpfinal}). When $\alpha=\pi/4$ the magnitude of the jumps can be estimated as $\epsilon^{1/2}Vl^{-1}_{\rm inh}$,
where $V$ is the phase speed of the incoming wave and $l_{\rm inh}$ is the characteristic scale of inhomogeneity. For typical
conditions in the solar upper chromosphere and corona the magnitude of the jumps in the derivatives of mean shear flow velocity
would be of the order of $10{\rm s}^{-1}$. From this, and the simple flow profiles given by Eqs. (\ref{eq:simpleprofilepara}) and (\ref{eq:simpleprofileperp}), we predict a mean flow outside the dissipative layer (generated by resonant absorption) with an amplitude
of the order of $10{\rm kms}^{-1}$ in the solar upper chromosphere and corona.

The magnitude of the jumps in the derivatives of the mean flow were found to depend on the phase speed ($V$), the characteristic scale of inhomogeneity ($l_{\rm inh}$) and the dimensionless amplitude of oscillation ($\epsilon$) far away from the dissipative layer. Even though simple estimations allowed us to approximate the magnitude of the shear flows, the results
in the present paper should be used with caution. The present analysis has been carried out for magnetic configurations that are homogeneous
and infinite in the direction of wave propagation outside the dissipative layer. This situation can only take place in laboratory devices (such as tokamaks). In the solar atmosphere, magnetic configurations are bounded and/or inhomogeneous in the direction of wave propagation. These additional
boundary conditions may reduce (or even prevent) the generation of mean shear flows by resonant absorption.

It is worth mentioning that when an \alf resonance is present there may exist a slow resonance coupled to it. If this happens,
both resonances will generated their own mean shear flows which could become superimposed. This may enhance the flow
through constructive superposition, but equally, they may interact destructively leaving a reduced overall combined generated mean shear flow.
In other words, it is possible to see a more efficient coupled dissipative layer create more heating, but actually generate a smaller amplitude mean shear flow compared with two separate dissipative layers.

If the boundary conditions allow the generation of a mean shear flows by resonant absorption in the solar atmosphere,
we would expect to see these flows superimposed on existing large scale motions. It is speculated that these mean shear flows (if set up) superimposed on existing bulk motion of the plasma would produce turbulent motions and create a Kelvin--Helmholtz instability (KHI) along the dissipative layer. A KHI instability may already exist due to the shear velocity field of \alf waves as was shown by Terradas \emph{et al.}\citet{terradas2008}. The generation of mean shear flow would survive the instability due to the velocity shear and, hence, will still contribute to the bulk motion causing larger shear, and perhaps creating a secondary instability. Either way, the KHI would distort the dissipative layer, locally enhancing the dissipation in (due to enhanced transport coefficients), which in turn would produce greater amplitude mean shear flows. This process of feedback would continue until either; (i) the wave is dissipated, or (ii) the turbulence destroys the dissipative layer. Either way, the turbulence would allow greater absorption and achieve heating over
a greater area (here the development of turbulent eddies can be seen as a means of heat transport).

The resonant absorption itself may not produce the heating we see in the solar corona, but we believe it is likely
that these dissipative layers are created and KHI are formed. The KHI might be able to trigger small scale reconnection events (nanoflaring) if the mean shear flows produced by resonant absorption become strong enough. However, further study and observations are required before we can make more definitive investigations.

\section*{ACKNOWLEDGEMENTS}

The authors wish to thank M.~S. Ruderman for helpful comments and discussions.
C.~T.~M. Clack would like to thank STFC (Science and Technology Facilities
Council) for the financial support provided. I. Ballai acknowledges
the financial support by NFS Hungary (OTKA, K67746) and The National
University Research Council Romania (CNCSIS-PN-II/531/2007).

\section*{APPENDIX: THE CALCULATION OF THE INTEGRAL OF THE $F$ FUNCTION}

In this Appendix, we will derive Eq. (\ref{eq:fcalculated}). We know the form of $F(\sigma)$ from
Eq. (\ref{eq:functionF}), so we can rewrite Eq. (\ref{eq:fcalculated}) as
\begin{multline}
\int^{\infty}_{-\infty} \left|\frac{dF}{d\sigma}\right|^2\mbox{ }d\sigma
=\int^{\infty}_{-\infty}d\sigma\left(\int^{\infty}_{0}\widetilde{\phi}e^{i\widetilde{\phi}\sigma-\widetilde{\phi}^{3}/3}\mbox{ }d\widetilde{\phi}\right.\\
\left.\times\mbox{ }\int^{\infty}_{0}\widetilde{\lambda}e^{-i\widetilde{\lambda}\sigma-\widetilde{\lambda}^3/3}\mbox{ }d\widetilde{\lambda}\right),\tag{A1}\label{eq:A1}
\end{multline}
where the tilde denotes the inclusion of the ${\rm sgn}(\Delta_a)$, from this point onwards we drop the tilde notation. Here we
are calculating the absolute value of the derivative, and so we use complex conjugates.
If we change the order the integration we obtain
\begin{multline}
\int^{\infty}_{-\infty} \left|\frac{dF}{d\sigma}\right|^2\mbox{ }d\sigma
=\int^{\infty}_{0}\phi e^{-\phi^3/3}\mbox{ }d\phi\\
\times\int^{\infty}_{0}\lambda e^{-\lambda^3/3}\mbox{ }d\lambda\times
\int^{\infty}_{-\infty}e^{-i\sigma\left(\lambda-\phi\right)}\mbox{ }d\sigma.\tag{A2}\label{eq:A2}
\end{multline}

The third integral in Eq. (\ref{eq:A2}) is the definition of the delta function, $\delta(\lambda-\phi)$, so we can
write Eq. (\ref{eq:A2}) as
\begin{multline}
\int^{\infty}_{-\infty} \left|\frac{dF}{d\sigma}\right|^2\mbox{ }d\sigma =
\int^{\infty}_{0}\phi e^{-\phi^3/3}\mbox{ }d\phi\\
\times\int^{\infty}_{0}\lambda e^{-\lambda^3/3}\mbox{ }\cdot 2\pi\delta(\lambda-\phi)d\lambda.\tag{A3}\label{eq:A3}
\end{multline}
Since the integral of the delta function is always unity, we arrive at
\begin{align}
\int^{\infty}_{-\infty} \left|\frac{dF}{d\sigma}\right|^2\mbox{ }d\sigma &=2\pi\int^{\infty}_{0}\phi^2 e^{-2\phi^3/3}\mbox{ }d\phi\nonumber\\
&=-\pi e^{-2\phi^3/3}\Big|^{\infty}_{0}=\pi.\tag{A4}\label{eq:A4}
\end{align}

\end{document}